\documentclass[aps,prb,twocolumn,showpacs,superscriptaddress]{revtex4-1}
\usepackage{graphicx,bm,amsmath,dcolumn,amssymb,color}
\graphicspath{{figures}}
\usepackage{longtable,geometry,hyperref}

\begin{document}
\title{Surface criticality of antiferromagnetic Potts model}
\author{Li-Ru Zhang}
\affiliation{School of Science and Engineering of Mathematics and Physics, Anhui University of Technology, Maanshan, Anhui 243002, China }

\author{Chengxiang Ding}
\email{dingcx@ahut.edu.cn}
\affiliation{School of Science and Engineering of Mathematics and Physics, Anhui University of Technology, Maanshan, Anhui 243002, China }

\author{Youjin Deng}
\affiliation{Department of Modern Physics, University of Science and Technology of China, Hefei, 230027, China}
\affiliation{MinJiang Collaborative Center for Theoretical Physics, College of Physics and Electronic Information Engineering,
Minjiang University, Fuzhou 350108, China}

\author{Long Zhang}
\email{longzhang@ucas.ac.cn}
\affiliation{Kavli Institute for Theoretical Sciences and CAS Center for Excellence in Topological Quantum Computation, University of Chinese Academy of Sciences, Beijing 100190, China}

\date{\today}

\begin{abstract}
 We study the three-state antiferromagnetic Potts model on the simple-cubic lattice, paying attention to the surface critical behaviors.
 When the nearest neighboring interactions of the surface is tuned, we obtain a phase diagram similar to the XY model,
 owing to the emergent O(2) symmetry of the bulk critical point.
 For the ordinary transition, we get $y_{h1}=0.780(3)$, $\eta_\parallel=1.44(1)$, and $\eta_\perp=0.736(6)$;
 for the special transition,
 we get $y_s=0.59(1)$, $y_{h1}=1.693(2)$, $\eta_\parallel=-0.391(4)$, and $\eta_\perp=-0.179(5)$; in the extraordinary-log phase, the surface correlation function $C_\parallel(r)$ decays logarithmically, with decaying exponent $q=0.60(2)$, however, the correlation $C_\perp(r)$ still decays algebraically, with critical exponent $\eta_\perp=-0.442(5)$.
	If the ferromagnetic next nearest neighboring surface interactions are added, we find two transition points,
	the first one is a special point between the ordinary phase and the extraordinary-log phase,
	the second one is a transition between the extraordinary-log phase and the $Z_6$ symmetry-breaking phase, with critical exponent $y_{\rm s}=0.41(2)$.
    The scaling behaviors of the second transition is very interesting,
    the surface spin correlation function $C_\parallel(r)$ and the surface squared staggered magnetization at this point decays logarithmically,
    with exponent $q=0.37(1)$; however, the surface structure factor with the smallest wave vector and the correlation function $C_\perp(r)$ satisfy power-law decaying, with critical exponents $\eta_\parallel=-0.69(1)$ and $\eta_\perp=-0.37(1)$, respectively.
\end{abstract}

\pacs{03.67.Bg, 03.65.Ud, 05.30.Rt}
\maketitle

\section{Introduction}
Phase transition and critical phenomena are hot topics in the research of condensed matter and statistical physics.
At the critical point of a continuous transition, the system exhibits a variety of singular behaviors characterized
by algebraically decaying of correlation functions. Such type of behaviors also appear on the surface of the system and can be different
to that of the bulk ones\cite{binder1974,binder1983,surf1,surf2,surf3}. Depending on the strength of the surface interactions, the surface critical behavior can be classified as
``ordinary transition", ``special transition", and ``extraordinary transition".  Typical examples can be found in the classical O($n$) models\cite{ising3Dsp,youjinOnsf,youjinO4sf}.
Generally, the ordinary transition can be found without tuning the surface interactions $J_{\rm s}$, i.e., it is the same as the bulk ones,
if it is tuned (strengthened), a phase transition may be found, which is the special point $J_{\rm sc}$, and the phase with $J_{\rm s}>J_{\rm sc}$ is the extraordinary phase. It is clear that the extraordinary phases of Ising model are ordered, however,
whether there is a special transition for the surface of the Heisenberg model and what the extraordinary phase is once controversial\cite{O3UC,O3sf2000,youjinOnsf}. Similar problems also exist in XY model.
Until recently, the research interest in surface critical phenomena was renewed by the exotic surface critical behavior in the quantum spin models\cite{Long2017,Ding2018,Weber2018,Weber2019}. A recent theoretical study pointed out that the extraordinary phase on the surface of the Heisenberg model may be an
``extraordinary-log" phase characterized by logarithmically decaying of the surface correlation function\cite{Metlitski2020}, which has been verified numerically\cite{O3sp}; similar behavior is also found in the XY model\cite{XYlog}.

\begin{figure}[htpb]
\includegraphics[width=0.85\columnwidth]{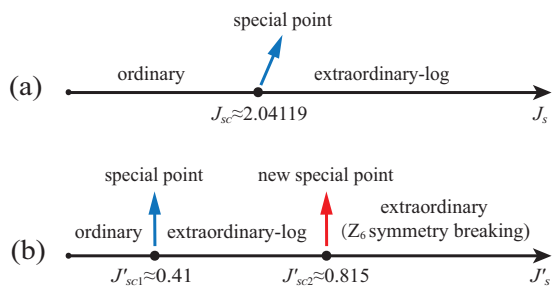}
\caption{Surface phase diagrams of the antiferromagnetic Potts model (\ref{Ham}): (a) the phase diagram with NNN interactions $J^\prime_{\rm s}=0$; (b) the phase diagram with NN interactions $J_{\rm s}=1$.}
\label{PD}
\end{figure}

The above results show that the surface critical phenomenon has many interesting characteristics, which are worth studying, and the related research are continuing\cite{Zhu2021,Weber2021,Toldin2021,Ding2021,Yu2021,Zhu2021-2,Jian2021,Max2111,Max2111a}.
In this paper, we study the surface critical properties of the three-state antiferromagnetic Potts model on a simple-cubic lattice\cite{AF3-1980,wsk}.
Due to the $Z_3$ symmetry of the spins and the permutation symmetry of two sets of sublattices, the ground state of the system breaks the $Z_6$ symmetry, however, at the bulk critical point, the symmetry of the order parameter is O(2), which is called ``emergent O(2) symmetry"\cite{Ding2014,Ding2016}, and the corresponding universality class of the phase transition is the same as the XY model.
At such ``emergent O(2)" bulk critical point, we explore the surface critical behaviors by
tuning the nearest neighboring (NN) surface interactions and the next nearest neighboring (NNN) surface interactions, the two phase diagrams we obtained are shown in Fig. \ref{PD}. In the phase diagram (a),
the extraordinary-log phase and the special point are obtained, which is very similar to the XY model\cite{youjinOnsf,XYlog}; we can also see that
the surface can not be ordered only by increasing the strength of the surface NN interactions,
this is closely related to the fact that the two-dimensional antiferromagnetic Potts model
is not ordered even the temperature is down to zero\cite{Ding2016}.
However, NNN ferromagnetic interactions can induce $Z_6$ symmetry breaking phase at finite temperature for the two-dimensional antiferromagnetic Potts model\cite{sqNijs},
this inspires us to add NNN ferromagnetic interactions to the surface of the three-dimensional antiferromagnetic Potts model; the results for this case are summarized in phase diagram (b), where a new special point with very interesting scaling behaviors is found.

The paper is arranged as follows: In Sec. \ref{model}, we introduce the model and method; in Sec. \ref{results}, we present the numerical results, including the refined bulk critical point, the critical behaviors about the ordinary transition, the extraordinary-log phase, and the two special points; we conclude our paper in Sec. \ref{con}.

\section{Model and Method}
\label{model}
The antiferromagnetic Potts model we studied is defined on the simple-cubic lattice
\begin{eqnarray}
	\mathcal{H}=J\sum\limits_{\langle i,j\rangle}\delta_{\sigma_i,\sigma_j}
	+J_s\sum\limits_{{\langle i,j\rangle}^s}\delta_{\sigma_i,\sigma_j}
	-J_s^\prime\sum\limits_{{\langle\langle i,j\rangle\rangle}^\prime}\delta_{\sigma_i,\sigma_j},\label{Ham}
\end{eqnarray}
$J$, $J_s$ and $J_s^\prime$ are the strengths of the NN bulk interactions, the NN surface interactions,
and the NNN surface interactions, respectively, noting that the NNN surface interactions are ferromagnetic.
The Potts spin $\sigma_i=1,2,3$, which can be mapped to a unit vector in the plane
\begin{eqnarray}
\vec{\sigma}_i=(\cos\theta_i,\sin\theta_i),
\end{eqnarray}
with $\theta_i=2\pi\sigma_i/3$.
Such mapping shows the symmetry of the spin, the variables we studied are based on such vectors.

The Monte Carlo algorithm we adopt is a combination of the local update (Metroplis algorithm) and global update (cluster algorithms)\cite{wsk}.
For the case $J^\prime_s\ne0$, the system has both antiferromagnetic and ferromagnetic interactions, the global update is a mixture of the Swendsen-Wang\cite{SW} and WSK algorithm\cite{wsk},
The high efficiency of the algorithm enables us to perform simulations for the systems with sizes up to $L=128$. In the simulations, the periodic boundary condition is applied
along the $x$ and $y$ directions, while open boundary condition is applied along the $z$ direction.

The bulk variables we sampled include the squared (staggered) magnetization $m_{\rm s}^2$, the magnetic susceptibility $\chi_s$, and the Binder Ratio $Q_{\rm s}$,
which are defined as
\begin{eqnarray}
	m_{\rm s}^2&=&\langle \mathcal{M}_{\rm s}^2 \rangle,  \\
\chi_{\rm s}&=&N(\langle\mathcal{M}_{\rm s}^2\rangle-\langle|\mathcal{M}_{\rm s}|\rangle^2),\\
Q&=&\frac{\langle \mathcal{M}_{\rm s}^2 \rangle^2}{\langle \mathcal{M}_{\rm s}^4\rangle},
\end{eqnarray}
where $T$ is the temperature and $\mathcal{M}_{\rm s}$ is defined as
\begin{eqnarray}
\mathcal{M}_{\rm s}&=&\frac{1}{N}\sum\limits_{\vec{R}}(-1)^{x+y+z} \vec{\sigma}_{\vec{R}} \label{Ms}.
\end{eqnarray}
Here $\vec{R}=(x,y,z)$ is the coordination, $N=L^3$ is the number of sites of the lattice.

We also sample the correlation function $C(r)$ and correlation length $\xi$
\begin{eqnarray}
&&C(r)=\langle\vec{\sigma}_i\cdot\vec{\sigma}_{i+r}\rangle,\label{Cr}\\
&&\xi=\frac{(m^2_{\rm s}/F-1)^{1/2}}{2\sqrt{\sum\limits_{i=1}^d \sin ^2(\frac{k_i}{2})}} \; ,
\end{eqnarray}
where $\vec{k}$ is the ``smallest wavevector" along the $x$ direction--i.e., $\vec{k} \equiv (2 \pi/L, 0, 0)$;
the ``structure factor" $F$ is define as
\begin{eqnarray}
F=\frac{1}{N^2}\Big\langle\big|\sum\limits_{\vec{R}}(-1)^{x+y+z}e^{i\vec{k} \cdot\vec{R}}\vec{\sigma}_{\vec{R}}\big|^2\Big\rangle. \label{F}
\end{eqnarray}
Generally, in a critical phase, correlation ratio $\xi/L$ assumes a universal nonzero value in the thermodynamic limit $L \rightarrow \infty$;
in a disordered phase, correlation length $\xi$ is finite and $\xi/L$ drops to zero,
while in an ordered phase, $\xi/L$ diverges quickly since the structure factor $F$ vanishes rapidly.
Therefore, like the Binder Ratio $Q_{\rm s}$, $\xi/L$ is also very useful in locating the critical point of phase transition.

The definition of the surface variables are very similar to the bulk ones, but the spins in Eqs. (\ref{Ms}), (\ref{Cr}), and (\ref{F}) should be restricted to the surface ones. For clarity, we add a subscript $1$ to the surface variables, i.e., the surface squared magnetization,
the surface magnetic susceptibility, the surface Binder Ratio, the surface structure factor, and the surface correlation length are written as $m_{\rm s1}$, $\chi_{\rm s1}$, $Q_{s1}$, $F_1$, and $\xi_1$, respectively, except the surface correlation functions, which are written as $C_\parallel(r)$ and $C_\perp(r)$; both $C_\parallel(r)$ and $C_\perp(r)$ are calculated as Eq. (\ref{Cr}), for $C_\parallel(r)$, both site $i$ and $i+r$ are on the surface; for $C_\perp(r)$, site $i$ is on the surface but site $i+r$ is in the bulk, and the line from $i$ to $i+r$ is perpendicular to the surface.

\section{Results}
\label{results}
\subsection{Refine the bulk critical point}
The simple-cubic antiferromagnetic Potts model has ever been studied by Monte Carlo simulations in Refs. \onlinecite{wsk} and \onlinecite{PottsWL},
 whose critical exponents coincide with the XY universality class, however, the accuracy is not very high.
 A more accurate numerical results about the critical exponents can
be found in Ref. \onlinecite{Ding2016}, while the interactions studied there are antiferromagnetic along the $x$ and $y$ directions
but ferromagnetic along the $z$ direction, such mixed interactions does not change the universality class of the phase transition but lead to different
(bulk) critical point. In current paper, before studying the surface critical behaviors, we refine the buk critical point
for the simple-cubic antiferromagentic model (with full antiferromagnetic interactions).

\begin{figure}[htpb]
\includegraphics[width=1.0\columnwidth]{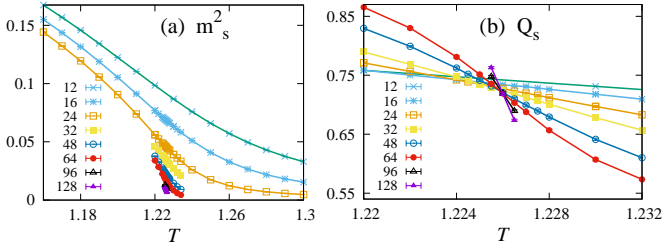}
\caption{Squared magnetization $m^2_{\rm s}$ and Binder Ratio  $Q_{\rm s}$ of the simple-cubic antiferromagnetic Potts model (\ref{Ham}).}
\label{qbulk}
\end{figure}
As shown in Fig. \ref{qbulk}, the bulk transition is clearly indicated by the bulk squared magnetization $m^2_{\rm s}$ and the Binder Ratio $Q_{s}$.
It can also be shown by the bulk correlation ratio $\xi/L$, which is not shown here.
In the vicinity of the critical point, $Q_{\rm s}$ and $\xi/L$ satisfy the finite-size scaling (FSS)\cite{fss1,fss2} formula
\begin{eqnarray}
	Q=a_0+\sum\limits_{k=1}^{k_{\rm max}}a_k(T-T_c)^kL^{ky_t}+bL^{y_1} \label{QFSS}
\end{eqnarray}
where $Q$ is $Q_{\rm s}$ or $\xi/L$, and $T_c$ is the critical point; $y_t>0$ is the thermal critical exponent, $y_1<0$ is the correction-to-scaling exponent,  $a_0$, $a_k$, and $b$ are unknown parameters.

The data fitting of $Q_{\rm s}$, with fixed $y_1=-1$, gives $T_c=1.22602(2)$ and $y_t=1.483(10)$;
the fitting from $\xi/L$ gives $T_c=1.22603(1)$ and $y_t=1.490(3)$;
here we chose the better one $T_c=1.22603(1)$ to be our final result, which will be used as the starting point of the study of surface critical behaviors in the next subsections. We can also see that the value of the critical exponent $y_t$ is consistent with that of the XY universality class\cite{Onloop,Lv2019}.

\subsection{Ordinary transition}
Exactly at the bulk critical point $T_c^{\rm bulk}=1.22603$, we perform simulations with open boundary condition along the $z$ direction and periodic boundary conditions along the $x$ and $y$ directions, the surface interactions are set as $J_{\rm s}=J=1$ and $J^\prime_{\rm s}=0$.
 In this case, the surface critical behaviors originates from the bulk correlations, which is called the
``ordinary transition".  The data of the surface squared magnetization $m^2_{\rm s1}$, the surface correlation function $C_{\parallel}$, the surface structure factor $F_1$, and the surface correlation $C_\perp$ are fit according to the scaling formulas
  \begin{eqnarray}
	&&m^2_{s1}L^2=c+L^{2y_{\rm h1}-2}(a+bL^{y_1}), \label{msurf_stag}\\
	&&C_\parallel(L/2)=L^{-1-\eta_\parallel}(a+bL^{y_1}),\label{Corp}\\
    &&F_1L^2=c+L^{2y_{\rm h1}-2}(a+bL^{y_1}),\label{F1}\\
    &&C_\perp(L/2)=L^{-1-\eta_\perp}(a+bL^{y_1}), \label{Corv}
\end{eqnarray}
where $y_{\rm h1}$, $\eta_\parallel$, and $\eta_\perp$ are critical exponent; $y_1<0$ is the correction-to-scaling exponents;
$a$, $b$, and $c$ are unknown parameters, with $c$ the analytical part of $m^2_{\rm s1}L^2$ or $F_1L^2$,
originating from the contribution of short-term correlations.
Although mathematically Eqs. (\ref{msurf_stag}) and (\ref{F1}) are equivalent to
\begin{eqnarray}
&&m^2_{s1}=c\cdot L^{-2}+L^{2y_{\rm h1}-4}(a+bL^{y_1}),\label{msurf_stag1}\\
&&F_1=c\cdot L^{-2}+L^{2y_{\rm h1}-4}(a+bL^{y_1})\label{F11},
\end{eqnarray}
respectively, technically, in the fitting of $m^2_{\rm s1}$, the RHS of (\ref{msurf_stag1}) or (\ref{F11}) will be dominated by the first term and the fitting
result of $y_{\rm h1}$ is often unreliable if the value of $y_{\rm h1}$ is smaller than $1$, which is exactly the case in ordinary transition.

The data fitting, with $y_1=-1$, gives $y_{\rm h1}=0.780(3)$ (from $m^2_{\rm s1}$), $y_{\rm h1}=0.782(4)$ (from $F_1$),
$\eta_{\parallel}=1.44(1)$, and $\eta_{\perp}=0.736(6)$,  these values coincide with those of the XY model\cite{youjinOnsf} and satisfy the scaling laws
\begin{eqnarray}
	&&\eta_\parallel=d-2y_{h1},\label{SL1}\\
    &&2\eta_\perp=\eta_\parallel+\eta,\label{SL2}
\end{eqnarray}
with $d=3$ the dimension of the system and $\eta=0.0385$ the exponent of the bulk correlation.

\subsection{Special transition and extraordinary phase by tuning the NN surface interaction $J_s$}
\begin{figure}[htpb]
\includegraphics[width=1\columnwidth]{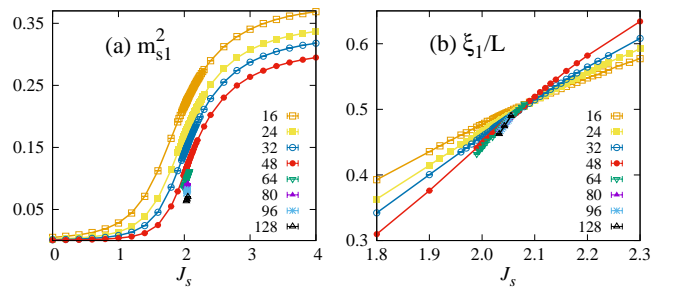}
\caption{Surface squared magnetization $m^2_{\rm s1}$ and correlation ratio $\xi_1/L$ of the antiferromagnetic Potts model (\ref{Ham}), with $T=T_c^{\rm bulk}=1.22603$ and the NNN surface interaction $J^\prime_{\rm s}=0$.}
\label{msxi}
\end{figure}
Exactly at the bulk critical point $T_c^{\rm bulk}=1.22603$, we tuning the NN surface interaction $J_s$ to see whether there is a special transition, this is confirmed by the behaviors of surface squared magnetization $m^2_{\rm s1}$ and surface correlation ratio $\xi_1/L$, as shown in Fig. \ref{msxi}.
This transition can also be detected by the Binder Ratio $Q_{\rm s1}$, which is not shown here.
The data of $\xi_1/L$ or $Q_{\rm s1}$ in the vicinity of the special point satisfy the scaling formula (\ref{QFSS}), with $T$ replaced by $J_{\rm s}$ and the critical exponent $y_t$ replaced by $y_{\rm s}$; the fitting, with $y_1=-1$, gives the critical point $J_{\rm sc}=2.04119(40)$ and the critical exponent $y_{\rm s}=0.59(1)$. We can see that the value of $y_s$ is consistent with that of the XY model\cite{youjinOnsf}.

Exactly at the special transition point $J_{\rm sc}=2.04119$, we investigate the scaling behaviors of the surface squared magnetization $m^2_{\rm s1}$, the surface correlation function $C_{\parallel}$, the surface structure factor $F_1$, and the surface correlation $C_\perp$,
they also satisfy the FSS formulas (\ref{msurf_stag}), (\ref{Corp}), (\ref{F1}), and (\ref{Corv}), respectively.
  Alternatively, $m^2_{\rm s1}$ and $F_1$ can also be fit
by Eqs. (\ref{msurf_stag1}) and (\ref{F11}) respectively, because in current case $y_{\rm h1}>1.5$, the RHS of the equation is dominated by the second term,
while the first one behaves like a correction-to-scaling term.
The data fitting gives $y_{\rm h1}=1.693(2)$ (from $m^2_{\rm s1}$), $y_{\rm h1}=1.694(4)$ (from $F_1$), $\eta_{\parallel}=-0.391(4)$, and $\eta_\perp=-0.179(5)$, these results coincide with those of the XY model and satisfy the scaling laws (\ref{SL1}) and (\ref{SL2}).

\begin{figure}[htpb]
\includegraphics[width=0.9\columnwidth]{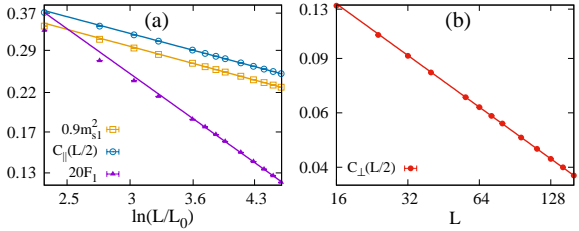}
\caption{(a) Log-log plot of $m^2_{\rm s1}$, $C_\parallel(L/2)$, and $F_1$ versus $\ln(L/L_0)$, for the antiferromagnetic Potts model (\ref{Ham}) with $T=T_c^{\rm bulk}=1.22603$, $J_{\rm s}=5$, and $J^\prime_{\rm s}=0$; $L_0=1.54$ is taken for all the three variables. (b) Log-log plot of the surface correlation $C_\perp(L/2)$.}
\label{js5}
\end{figure}
For $J_{\rm s}>J_{\rm sc}$, we find that the system is in the extraordinary-log phase, characterizing by the logarithmically decaying of correlation function $C_\parallel(r)$ and related variables\cite{Metlitski2020,O3sp,XYlog}.
 Fig. \ref{js5}(a) shows a typical case of $J_s=5$,  and the data can be fit according to
\begin{eqnarray}
    &&m^2_{\rm s1}=a\cdot[\ln(L/L_0)]^{-q},\label{logm}\\
    &&C_\parallel(L/2)=a\cdot[\ln(L/L_0)]^{-q},\label{logCp}\\
    &&F_1=a\cdot[\ln(L/L_0)]^{-q^\prime}\label{logF1},
\end{eqnarray}
where $q$ and $q^\prime$ are the critical exponents, $a$ and $L_0$ are nonuniversal parameters.
From the fitting of $C_\parallel$, we get $q=0.60(2)$ and $L_0=1.54(9)$;
from the fitting of $m^2_{\rm s1}$ and $F_1$, we get $q=0.61(1)$ and $q^\prime=1.61(1)$, respectively.
We can see that the values of $q$ and $q^\prime$ are consistent with those of the XY model\cite{XYlog} and satisfy the relation $q^\prime=1+q$.
It should be noted that in the fitting of $m^2_{\rm s1}$ and $F_1$, we have set the value of $L_0$ fixed to 1.54,
such trick is the same as that in Ref. \onlinecite{XYlog} for XY model.
For the surface correlation function $C_\perp(L)$, we find that it still decaying algebraically, as shown in Fig. \ref{js5}(b), therefore we fit it according to Eq. (\ref{Corv}) with $y_1=-1$, we get $\eta_\perp=-0.442(5)$.

We also study the case of $J_s=10$, and get $q=0.63(2)$, $q^\prime=1.64(2)$, $L_0=1.2(1)$, and $\eta_\perp=-0.444(7)$.
The values of $q$ and $q'$ coincide with the cases of $J_s=5$ and the XY model\cite{XYlog},
which means the critical behaviors of the extraordinary-log phase of the O(2) surface are universal.

In the low temperature, the order parameter of the antiferromagnetic Potts model on the simple-cubic lattice breaks the $Z_6$ symmetry\cite{Ding2016},
therefore one may ask whether the surface can be ordered with such symmetry breaking when the surface interactions are strengthened.
However, by performing the simulations with much larger $J_s$, we find that the surface is always in the extraordinary-log phase,
it can not be ordered by the strengthen of $J_s$. This may be related to the
fact that in the limit of $J_s\rightarrow \infty$, the decoupled two dimensional antiferromagnetic Potts model can not be ordered even the temperature is zero\cite{pottssq}.
However, as we will show in the next subsection, the $Z_6$ symmetry breaking surface can be reached by adding the NNN ferromagnetic interactions $J^\prime_s$ in the surface.

\subsection{New special transition induced by next nearest interactions $J^\prime_s$}
\begin{figure}[htpb]
\includegraphics[width=1.0\columnwidth]{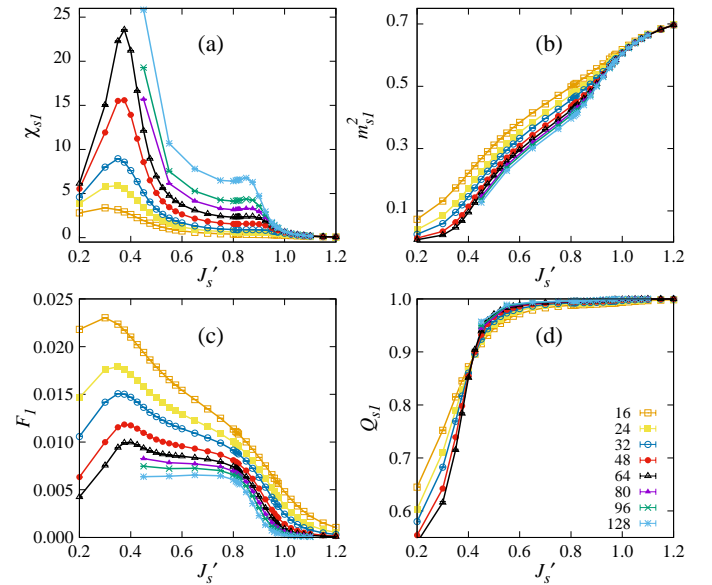}
\caption{Surface critical behaviors of the antiferromagnetic Potts model (\ref{Ham}), with $T=T_c=1.22603$ and $J_{\rm s}=1$:
(a) surface susceptibility $\chi_{\rm s1}$, (b) surface squared magnetization $m^2_{\rm s1}$, (c) surface structure factor $F_1$, (d) surface Binder Ratio $Q_{\rm s1}$. }
\label{twopeak}
\end{figure}

When the NNN ferromagnetic interactions $J^\prime_s$ is added to the surface, we can find two phase transitions, as shown by the behaviors of the surface susceptibility $\chi_{\rm s1}$ in Fig. \ref{twopeak}(a).
In the intermediate region $J^\prime_{\rm sc1}<J^\prime_{\rm s}<J^\prime_{\rm sc2}$, with $J^\prime_{\rm s1}\approx 0.41$ and $J^\prime_{\rm sc2}\approx 0.85$,
the scaling behaviors of $\chi_{\rm s1}$ is very different from that in region $J^\prime_{\rm s}<J^\prime_{\rm sc1}$ or $J^\prime_{\rm s}>J^\prime_{\rm sc2}$.
As we will shown later, in this region, the system is in the extraordinary-log phase.
Therefore, the peak of $\chi_{\rm s1}$ at $J^\prime_{\rm sc1}\approx0.41$  indicates a special transition point like that studied in above subsection.

\begin{figure}[htpb]
\includegraphics[width=1.0\columnwidth]{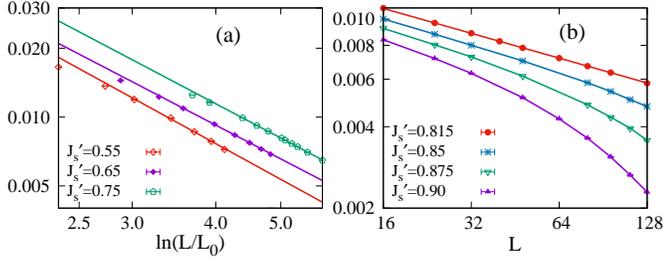}
\caption{(a) Log-log plot of $F_1$ versus $\ln(L/L_0)$, using the values of $L_0$ fit from $m^2_{s1}$, listed in Table \ref{newpoint}; (b) log-log plot of
$F_1$ versus $L$.}
\label{F1fit}
\end{figure}
The second transition at $J^\prime_{\rm sc2}$, which we call ``new special point", is revealed not only by the surface susceptibility $\chi_{s1}$, but also by the surface squared magnetization $m^2_{\rm s1}$ and surface structure factor $F_1$, as shown in Fig. \ref{twopeak}(b) and (c), respectively;
however, the signature of the Binder Ratio $Q_{\rm s1}$ for this transition is not obvious, as shown in Fig. \ref{twopeak}(d).

We can see that the second peak of $\chi_{\rm s1}$ diverges much slower than the first one, namely the singularity of such transition is very weak.
The transition point $J^\prime_{\rm sc2}\approx 0.85$ is obtained by the position of such peak of the largest system with size $L=128$;
however, considering the finite-size effect,
 the critical point in the thermodynamic limit should be a little smaller than this one.
 A more accurate estimation of the critical point can be obtained from the scaling behavior of the structure factor $F_1$, which satisfies the logarithmic decaying
 in the region $J^\prime_{\rm sc1}<J^\prime_{\rm s}<J^\prime_{\rm sc2}$
 but a power-law scaling at the transition point $J^\prime_{\rm sc2}=0.815$; this is demonstrated in Fig. \ref{F1fit}.

 In the region $J^\prime_{\rm sc1}<J^\prime_{\rm s}<J^\prime_{\rm sc2}$, we check the FFS behaviors of $F_1$ for several cases,
 the results of data fitting, according to Eq. (\ref{logF1}), are listed in Table. \ref{newpoint}.
 In this region, $m^2_{\rm s1}$ and $C_\parallel$ also satisfy the logarithmic decaying, the fitting results, according to Eq. (\ref{logm}) and (\ref{logCp}),
 are also listed in Table. \ref{newpoint}. These results claim that in this region the system is in an extraordinary-log phase;
 the critical exponents $q$ and $q^\prime$ in this phase are universal and satisfy the relation $q^\prime=1+q$, as that in XY model\cite{XYlog}.
 In this phase, the correlation $C_\perp$ still satisfy the power-law scaling formula (\ref{Corv}),
 the fitting results of $\eta_\perp$ for several cases are also listed in Table. \ref{newpoint}.

 An important point should be emphasized about the extraordinary-log phase is that the surface susceptibility $\chi_{\rm s1}$ is divergent in this phase,
 such property is very different from that of the ordinary phase or the extraordinary phase with long-range order. Technically, this property
 can help us to detect whether a surface is in an extraordinary-log phase. It should be noted that, currently, it not a trivial work to distinguish an
 extraordinary-log phase from a long-range ordered phase numerically without theoretical precognition, especially when the value of the long-range order is relatively small.
  \begin{table}[thbp]
\caption{Fitting results of $m^2_{\rm s1}$, $C_\parallel$, and $F_1$, where $m^2_{\rm s1}$ and $C_\parallel$ are fit according to Eqs. (\ref{logm}) and (\ref{logCp}), respectively; $F_1$ is fit according to (\ref{logF1}) when $J^\prime_{\rm s}<J^\prime_{sc2}$ but a power-law formula $aL^{-1-\eta_\parallel}$ when $J^\prime_{\rm s}<J^\prime_{sc2}$. In fitting of $C_\parallel$ and $m^2_{\rm s1}$, we let $L_0$ free to be fit, while in the fitting of $F_1$, we fix $L_0$ as that of $m^2_{\rm s1}$. }
\begin{tabular}{l |c c| c  c |c |c c}
\hline
  $J^\prime_{\rm s}$ &$q$ ($C_\parallel$) &$L_0$ & $q$ ($m^2_{\rm s1}$) &$L_0$& $q'$ &$\eta_\perp$ \\
  0.55  &            0.61(3) &2.0(7)    &0.59(3)  &2.63(9)  &1.56(4)  &-0.43(1) \\
  0.65  &            0.56(2) &1.36(10)  &0.60(2)  &0.89(5)  &1.59(1)  &-0.43(1)\\
  0.75  &            0.55(2) &0.90(16)  &0.57(1)  &0.44(9)  &1.57(1)  &-0.38(1) \\
  \hline
  $J^\prime_{\rm sc2}$ &$q$ ($C_\parallel$) &$L_0$ & $q$ ($m^2_{\rm s1}$) &$L_0$& $\eta_\parallel$ &$\eta_\perp$ \\
  0.815 &            0.37(1) &2.3(3)    &0.38(1)  &1.13(9) &$-0.69(1)$ & -0.37(1) \\
 \hline
\end{tabular}
\label{newpoint}
\end{table}

\begin{figure}[htpb]
\includegraphics[width=1.0\columnwidth]{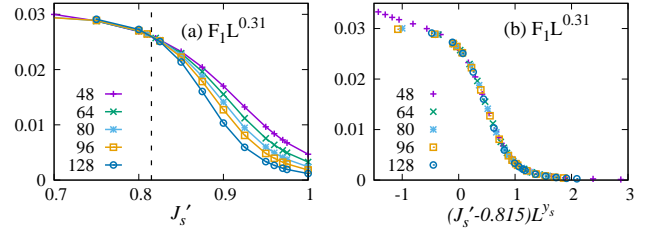}
\caption{(a) $F_1^{-1-\eta}=F_1^{0.31}$ versus $J^\prime_{\rm s}$; the dashed line is the critical point; (b) data collaps: $F_1L^{0.31}$ versus $(J^\prime_{\rm s}-0.815)L^{y_s}$, with $y_s=0.41$.}
\label{Fcollaps}
\end{figure}
 At the transition point $J^\prime_{sc2}=0.815$, $F_1$ satisfy a power-law formula $aL^{-1-\eta_\parallel}$, where $\eta_\parallel=-0.69(1)$.
The correlation function $C_\perp$ also satisfies the power law (\ref{Corv}), and the critical exponent $\eta_\perp=-0.37(1)$.
 However, the $m^2_{\rm s1}$ and $C_\parallel$ at this point still satisfy the logarithmically decaying formulas (\ref{logm}) and (\ref{logCp}),
 with exponent $q=0.30(1)$, as shown in Table \ref{newpoint}.

 In the range $J^\prime_{\rm s}>J^\prime_{\rm sc2}$, the surface is in a long-range ordered phase, the structure factor $F_1$ decays much faster,
 which satisfies neither a logarithmically decaying formula nor a simple pow-law formula, this is also demonstrated in Fig. \ref{F1fit}.

 In order to further explore the critical behaviors of the new special point, we plot $F_1L^{-1-\eta_\parallel}=F_1L^{0.31}$  versus $J^\prime_{\rm s}$ in Fig. \ref{Fcollaps}(a), we can see that $F_1L^{-1-\eta_\parallel}$ plays the role of a dimensionless variable like the Binder Ratio;
 Fig. \ref{Fcollaps}(b) is the data collaps plot of $F_1L^{0.31}$ versus
 $(J^\prime_{\rm s}-0.815)L^{y_{\rm s}}$, with critical exponent $y_{\rm s}=0.41$.
 More rigorously, one can fit the data of $F_1$ in the
 vicinity of $J^\prime_{\rm s}=0.815$  by the scaling formula
 \begin{eqnarray}
 F_1=L^{-1-\eta_\parallel}\big[a_0+\sum\limits_{k=1}^{k_{\rm max}}a_k(J^\prime_{\rm s}-J^\prime_{\rm sc2})^kL^{ky_s}+bL^{y_1}\big], \label{F1FSS}
 \end{eqnarray}
which gives $\eta_\parallel=-0.69(1)$, $J^\prime_{\rm sc2}=0.815(2)$, and $y_{\rm s}=0.41(2)$, these results are consistent with those applied in the data collaps in Fig. \ref{Fcollaps}(b),  thus give a self-consistent check.

 In summary, for the new special transition, it is mainly characterized by the the following critical behaviors,
 \begin{eqnarray}
 &&J^\prime_{\rm sc2}=0.815(2),\\
 &&y_s=0.41(2),\\
 &&F_1\sim aL^{-1-\eta_\parallel}, {\rm with} \ \eta_\parallel=-0.69(1),\\
 &&m^2_{\rm s1}\sim a\cdot [\ln(L/L_0)]^{-q}, {\rm with} \ q=0.37(1),\\
 &&C_\parallel \sim a\cdot [\ln(L/L_0)]^{-q}, {\rm with} \ q=0.38(1),\\
 &&C_\perp\sim aL^{-1-\eta_\perp}, {\rm with} \quad \eta=-0.37(1).
 \end{eqnarray}

 \begin{figure*}[htpb]
\includegraphics[width=1.6\columnwidth]{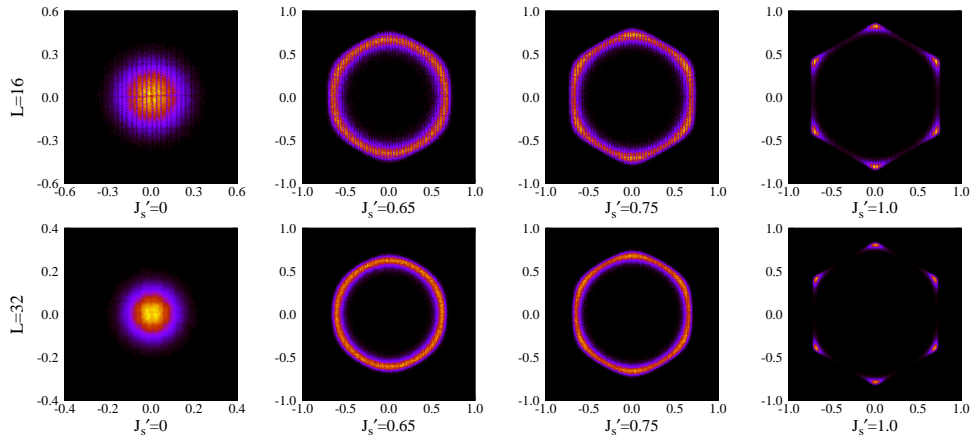}
\caption{Histograms of the surface staggered magnetization of the antiferromagnetic Potts model (\ref{Ham}); in all the cases, the strength of the surface NN interactions $J_{\rm s}=1$.}
\label{histo}
\end{figure*}
\subsection{Symmetries of the surface}
After studied the phase transitions, we give a short description about the symmetries of the model.
The three-state antiferromagnetic Potts model is well known for the emergent O(2) symmetry at the bulk critical point, although the ground state is $Z_6$ symmetry breaking, which is ordered by entropy.  It is an interesting question what the surface symmetry is.

As shown in Fig. \ref{histo},
in the ordinary phase ($J^\prime_{\rm s}=0$), the histogram visualizes the disorder properties of
the surface staggered magnetization, this confirms that the surface is disorder, and the critical behaviors are purely induced by the bulk criticality. For the extraordinary-log phase, we show two cases; for the case of $J^\prime_{\rm s}=0.65$, we can see that symmetry of the order parameter is O(2)
when the system size reaches $L=32$, although we can still find the shadow of the $Z_6$ symmetry when the system size is small ($L$=16); for the case of
$J^\prime_{\rm s}=0.75$, we can see that the $Z_6$ symmetry persists to show its effect when the system size is $L=32$, however, the effect is
relatively weak comparing to that of $L=16$, thus we infer that in the thermodynamic limit, the symmetry should be O(2). Such property is important,
it is also helpful for us to numerically distinguish the extraordinary-log phase from an ordered surface with small value of long-range order.

When $J^\prime_{\rm s}$ is large enough, the surface is ordered and the $Z_6$ symmetry is very obvious, this is also shown in Fig. \ref{histo} with $J^\prime_{\rm s}=1.0$.

\section{Conclusion and discussion}
\label{con}
In summary, we have studied the surface critical behaviors of the three-state antiferromanetic Potts model on the simple-cubic lattice,
we obtain a phase diagram similar to the XY model by tuning the NN surface interactions, the universality classes of the special transition and extraordinary-log phase are the same as the XY model. A long range-order surface which breaks the $Z_6$ symmetry and a new type of special transition between such ordered phase and the extraordinary-log phase is obtained by tuning the NNN ferromagnetic surface interactions.
At such new special transition point, the scaling behaviors are very interesting; the surface squared magnetization $m^2_{\rm s1}$ and the surface correlation function
$C_\parallel$ satisfy the logarithmic decaying, with exponent $q=0.37(1)$, but the structure factor $F_1$ and the correlation function $C_\perp$ still
satisfy the power-law decaying, with critical exponent $\eta_\parallel=-0.69(1)$ and $\eta_\perp=-0.37(1)$, respectively. We also visualized the symmetries of
different phases of the surface, where the extraordinary-log phase is shown to conserve O(2) symmetry.

We have to stress that all our conclusions about the new special transition are based on the assumption that it is a second-order phase transition, which does not rule out the possibility of other types of phase transitions, such as the Berezinskii-Kosterlitz-Thouless transition.

In classical XY model, although the symmetry of the bulk critical point is also O(2),
the surface can not be ordered because of Mermin-Wagner-Hohenberg theorem\cite{MW1966,H1967},
 therefore such new type of special transition can not appear in classical XY model;
discrete Hamiltonian symmetry and emergent O($n$) symmetry of bulk criticality seems a necessary condition for such new type of special transition,
such as the clock model\cite{clock6}.
In addition to the emergent O(2) critical point, other emergent O($n$) critical point have also been found
in systems with discrete symmetry of Hamiltonian\cite{Ding2016,Leonard2015}. It is obviously an important and interesting question to study
 such new special transition and related critical properties in these models.

\section{Acknowledgment}
We thank Jian-Ping Lv for valuable discussions, and Cenke Xu for the advice of visualizing the symmetries of the sufrace by histograms.
C.D. is supported by the National Science Foundation of China under Grants Numbers 11975024 and 62175001,
the Anhui Provincial Supporting Program for Excellent Young Talents
in Colleges and Universities under Grant Number gxyqZD2019023.
L.Z. is supported by the National Key
R\&D Program (2018YFA0305800), the National Natural
Science Foundation of China under Grants Numbers
11804337 and 12174387, CAS Strategic Priority Research
Program (XDB28000000) and CAS Youth Innovation
Promotion Association.
Y.D. is supported by the National Natural Science Foundation of China under Grant No. 11625522, the Science and Technology Committee
of Shanghai under grant No. 20DZ2210100, and the National Key R\&D Program of China under Grant No. 2018YFA0306501.

\end{document}